\documentclass[12pt]{article}
\usepackage{graphicx}
\usepackage{amsmath}
\usepackage{longtable}

\begin{document}

% 4 rows Astronomy Letters style:
\makeatletter
\renewcommand*{\@cite}[2]{{#2}}
\renewcommand*{\@biblabel}[1]{#1.\hfill}
\makeatother

\title{Dependence of Kinematics on the Age of Stars in the Solar Neighborhood}
\author{G.~A.~Gontcharov\thanks{E-mail: georgegontcharov@yahoo.com}}

\maketitle

Pulkovo Astronomical Observatory, Russian Academy of Sciences, Pul\-kov\-skoe sh. 65, St. Petersburg, 196140 Russia


The variations of kinematic parameters with age are considered for a sample of 15 402 thin-disk
O--F stars with accurate $\alpha$, $\delta$, $\mu$, and $\pi>3$ mas from the Hipparcos catalogue and
radial velocities from
the PCRV catalogue. The ages have been calculated from the positions of the stars on the Hertzsprung--Russell
diagram relative to the isochrones from the Padova database by taking into account the extinction
from the previously constructed 3D analytical model and extinction coefficient $R_V$ from the 3D map of its
variations. Smooth, mutually reconciled variations of the velocity dispersions $\sigma(U)$, $\sigma(V)$, $\sigma(W)$,
solar motion components $U_{\odot}$, $V_{\odot}$, $W_{\odot}$,
Ogorodnikov--Milne model parameters, Oort constants, and vertex
deviation $l_{xy}$ consistent with all of the extraneous results for which the stellar ages were determined have
been found. The velocity dispersion variations are well fitted by power laws the deviations from which are
explained by the influence of predominantly radial stellar streams: Sirius, Hyades, $\alpha$~Cet/Wolf~630, and
Hercules. The accuracy of determining the solar motion relative to the local standard of rest is shown to be
fundamentally limited due to these variations of stellar kinematics. The deviations of our results from those
of Dehnen and Binney (1998), the Geneva--Copenhagen survey of dwarfs, and the Besancon model of the
Galaxy are explained by the use of PCRV radial velocities with corrected systematic errors.

\newpage
\section*{INTRODUCTION}

Among the kinematic parameters of stars, the following
parameters are most informative and are traditionally
considered: the characteristics of the distribution
of stars in the phase space of velocity components
$U$, $V$, $W$ in the Galactic coordinate system,
their dispersions $\sigma(U)$, $\sigma(V)$, $\sigma(W)$, the parameters
of the linear Ogorodnikov--Milne model, the related
Oort constants $A$, $B$, $C$, $K$, and the components of
the peculiar solar motion (motion toward the apex)
relative to the local standard of rest (LSR). By the
latter we mean an imaginary star or a reference point
that is presently at the location of the Sun and moves
uniformly in a circular orbit in the real gravitational
potential of the Galaxy averaged over the Galactic
longitude.

The estimates of kinematic parameters for different
samples of stars are contradictory (see, e.g., Francis
and Anderson 2009). This forces one to search for not
just the age--velocity relation but their smooth, physically
justified and reconciled (for different parameters)
variations with stellar age.

Here, we calculate the components of the stellar
positions and motions in the Galactic coordinate system
(the positions in 6D space $XYZUVW$,
the $X$ axis is directed to the Galactic center, the $Y$ axis is
in the direction of Galactic rotation, and the $Z$ axis is
directed to the North Pole) for a sample of early-type
main-sequence (MS) stars. We independently estimate
the most probable ages of these stars, establish
the variations of kinematics with age, and compare
them with extraneous results.

\section*{COORDINATES AND VELOCITIES OF THIN-DISK STARS}

The Pulkovo Compilation of Radial Velocities
(PCRV, Gontcharov 2006) is currently the largest
source of barycentric radial velocities of stars ($V_r$)
that are fairly free from systematic errors. It contains
$V_r$ for 35 493 stars from the Hipparcos catalogue
(the first version -- ESA (1997); the new version -- van
Leeuwen (2007)). The distribution of these stars on
a Hertzsprung -- Russell (H--R) diagram, for example,
of the form ``color -- absolute magnitude'', corresponds
to the distribution of all Hipparcos stars. The median
accuracy of $V_r$ from the PCRV is 0.7 km s$^{-1}$; for
all stars, $V_r$ is more accurate than 5 km s$^{-1}$. The
PCRV includes $V_r$ from 203 catalogues. The two
largest present-day catalogues are among them: the
Geneva--Copenhagen survey of more than 14 000
stars mostly of types FV--GV near the Sun (GCS)
(Nordstr\"om et al. 2004; Holmberg et al. 2007,
2009) and the kinematic survey of more than 6000
KIII--MIII stars based on CORAVEL observations
(Famaey et al. 2005).

The desire to minimize the systematic errors of
the trigonometric parallaxes due to the Lutz--Kelker
and Malmquist biases (Perryman 2009, pp. 209--211)
forces us to limit the sample of stars in parallax: $\pi>3$
milliarcseconds (mas).

Having supplemented the data on $V_r$ and $\pi$ with
the coordinates $\alpha$ and $\delta$ and proper motions $\mu$ from the
new version of Hipparcos, we calculate the complete
set of $XYZUVW$ for 26 399 stars.

The accuracy of the components $\mu_{\alpha}\cos\delta$ and
$\mu_{\delta}$ for the overwhelming majority of sample stars is
higher than 3 mas yr$^{-1}$. At a distance of 333 pc,
this corresponds to an accuracy of 5 km s$^{-1}$ for the
velocity components. The accuracy of the Vr used
is also higher than 5 km s$^{-1}$. Thus, the sample is
homogeneous with regard to the accuracy of $U$, $V$, and $W$.

The dereddened color $(B_{T}-V_{T})_{0}$ was calculated
for each star:
\begin{equation}
\label{btvt0}
(B_{T}-V_{T})_{0}=(B_{T}-V_{T})-E(B_{T}-V_{T}),
\end{equation}
where the reddening $E(B_{T}-V_{T})=A_{V_{T}}/R_{V}=1.1A_{V}/R_{V}$
and the extinction $A_V$, in turn, was
calculated from the 3D analytical extinction model
(see Gontcharov 2009, 2012b) as a function of the
trigonometric distance $r=1/\pi$ and Galactic coordinates
$l$ and $b$, while the extinction coefficient $R_V$ was
calculated from the 3D map of its variations as a function
of the same coordinates (Gontcharov 2012a).
The absolute magnitude $M_{V_{T}}$ was calculated for each
star from the formula
\begin{equation}
\label{mvt}
M_{V_{T}}=V_{T}+5-5\lg~r-A_{V_{T}}.
\end{equation}

Although below we consider only early-type MS
stars ($(B_{T}-V_{T})_{0}<0.7$), thick-disk and halo stars
(for example, the hot and cool subdwarfs considered
by Gontcharov et al. (2011)) are also encountered
among them. Here, we consider only Galactic
thin-disk stars. We assigned 15 730 sample stars
with $(B_{T}-V_{T})_{0}<0.7$ to the thin disk, the thick
disk, or the halo in agreement with the criteria from
the Besancon model of the Galaxy (BMG) (Robin et al. 2003):
\begin{eqnarray*}
\label{bmg}
&\overline{U}-3\sigma_{U}<U<\overline{U}+3\sigma_{U}, \\
&\overline{V}-3\sigma_{V}<V<\overline{V}+3\sigma_{V}, \\
&\overline{W}-3\sigma_{W}<W<\overline{W}+3\sigma_{W},
\end{eqnarray*}
where $\overline{U}=-10$, $\overline{V}=-10+0.84e^{3.86(B_{T}-V_{T})_{0}}$ (this
dependence is justified below), and $\overline{W}=-7$
are the mean values of $U$, $V$, $W$ for all stellar
populations; $\sigma_{U}$, $\sigma_{V}$, $\sigma_{W}$ are the boundary
velocity dispersions for the three populations. We
assigned 15 402 stars with $\sigma_{U}=43.1$, $\sigma_{V}=(27.8^2+14.8^2)^{1/2}$, $\sigma_{W}=24$
km s$^{-1}$ to the thin disk,
214 stars with $\sigma_{U}=67$, $\sigma_{V}=(51^2+49^2)^{1/2}$, $\sigma_{W}=42$
km s$^{-1}$ to the thick disk, and the remaining
114 stars to the halo; $\sigma_{V}$ is the quadratic sum of the
dispersion proper (the first term) and the asymmetric
drift (the second term), a systematically different
Galactic rotation velocity for different populations. In
contrast to the BMG, we chose $\sigma_{W}=24$ instead of
17.5 km s$^{-1}$ for the thin disk and an asymmetric drift
of 49 instead of 53 km s$^{-1}$ for the thick disk. These
differences are attributable to the distribution of the
15 730 stars under consideration in $U$, $V$, and $W$
shown in Fig. 1: the local minima and changes in the
smooth behavior of the curve point to the population
boundaries.

The thick-disk and halo stars account for 1.4\%
and 0.7\% of the sample, respectively. In the BMG,
these ratios are 3.3\% and 0.02\% for stars of all ages,
with the exception of white dwarfs. Since there should
be fewer thick-disk and halo stars among the O--F
stars and among the G--M ones, their fraction in the
sample under consideration appears plausible.

\section*{CALCULATING THE AGES}

The age of a star is usually estimated from its
position on the H--R diagram relative to the theoretical
isochrones. Since the isochrones differ significantly
for stars of different metallicities in many
regions of the H--R diagram, it is desirable to know
the stellar metallicity in advance and to consider the
set of isochrones for this metallicity. The ages of
stars in the GCS were estimated precisely in this way.
The metallicity was determined from narrow-band
Str\"omgren photometry. Unfortunately, Str\"omgren
photometry and the calibrations allowing the ages to
be estimated from it are available mostly for dwarfs
older than 2 Gyr, whichwere investigated in the GCS.
There are almost no stars younger than 1 Gyr in the
GCS. Since the sought-for variations of kinematics
with age can be smoothed out and disappear, it
is more fruitful to search for them in young stars.
Therefore, instead of Str\"omgren photometry, we use
Hipparcos/Tycho-2 photometry (H\o g et al. 2000) in
the broad $B_T$ and $V_T$ bands to estimate the ages and
consider not the individual ages but the mean ones for
the sets of stars. Naturally, in order that the concepts
of mean age and mean metallicity be meaningful, we
will consider only the regions of the H--R diagram
where the scatter of these quantities is small. In addition,
we will exclude the regions of the H--R diagram
where the isochrones for different metallicities are far
from each other from consideration.

Figure 2 shows the correlation between dereddened
color $(B_{T}-V_{T})_{0}$ and age from the GCS for the
8795 thin-disk stars under consideration. The solid
curve indicates an exponential fit to these data for age
$T$ (Gyr):
\begin{equation}
\label{exp1}
T=0.41e^{3.9(B_{T}-V_{T})_{0}}.
\end{equation}
The relative accuracy of the GCS ages lies predominantly
within the range 40--70\% and causes a large
scatter of ages seen in the figure and even the presence
of ages older than the presumed age of the Universe.
The scatter of true ages is apparently smaller
and using the mean age is justified in the entire range
$(B_{T}-V_{T})_{0}<0.65^m$ (in our calculations, we also used
stars with $0.65^m<(B_{T}-V_{T})_{0}<0.7^m$, but their kinematics
is not considered). This limitation implies the
selection of O--G2 stars (hotter and mostly younger
than the Sun).

The positions of the stars under consideration
on the H--R diagram are shown in Fig. 3 before
($(B_T-V_T)$ -- $M_{V_{T}}$) and after ($(B_T-V_T)_0$ -- $M_{V_{T}}$)
dereddening. The cross indicates typical errors for an
individual star: $\sigma(B_T-V_T)=0.02$ and $\sigma(M_{V_{T}})=0.5^{m}$.
The curves indicate the isochrones of solar metallicity
($FeH=0$) stars for ages of 0.001, 0.1, 0.2,
0.4, 1, 2, 3, and 4 Gyr. The 0.001-Gyr isochrone may
be considered the zero-age main sequence (ZAMS).
Here and below, the isochrones were taken from
the Padova database (http://stev.oapd.inaf.it/cmd;
Bertelli et al. 2008; Marigo et al. 2008).

We see from Fig. 3 that the applied 3D extinction
model successfully placed the stars into the expected
region of the diagram, predominantly between the
ZAMS and the loops of isochrones implying the beginning
of the subgiant stage (hydrogen burning in
the shell above an inert helium core). We also see that
there are almost no giants and low-metallicity stars
(below the ZAMS) among the thin-disk stars in the
range $(B_{T}-V_{T})_{0}<0.7^m$. The latter fact can also be
seen in the GCS: the mean metallicity for dwarfs with
an age of 5 Gyr from the GCS is $\overline{FeH}\approx-0.15$ (while
the universally accepted boundaries between the thin
disk, the thick disk, and the halo are $FeH\approx-0.4$ and
$FeH\approx-1.3$).

We see from Figs. 2 and 3 that there is a strong
correlation between age, dereddened color, and metallicity
for early-type MS stars pointed out by the GCS
authors. According to the GCS, the age dependence
of FeH can be represented with an adequate accuracy
by the linear trend $\overline{FeH}=0.01-0.034\cdot T$, or, given
dependence (3),
\begin{equation}
\label{feh}
\overline{FeH}=0.01-1.394e^{3.9(B_{T}-V_{T})_{0}}.
\end{equation}
Using this dependence, we find that only for stars
older than 2 Gyr are the $\overline{FeH}$ isochrones offset
from the $FeH=0$ isochrones on the H--R diagram
by more than the accuracy of the photometry
used. Even for the oldest stars under consideration,
the shift of the $\overline{FeH}$ isochrone from the $FeH=0$
isochrone is ($\Delta((B_{T}-V_{T})_{0})<0.06^m$, $\Delta(M_{V_{T}})<0.2^{m}$),
i.e., less than the above mean stellar position
error on the H--R diagram due to the parallax error.
Therefore, the change in $FeH$ during the last 5 Gyr
has almost no effect on the stellar ages calculated
below.

The Padova database allows us to interpolate the
isochrones with a high accuracy and to unambiguously
select the isochrone for each star under consideration
by taking into account the metallicity adopted
for the star using Eq. (4). Thus, we calculated the
individual ages for all stars in the range $(B_{T}-V_{T})_{0}<0.7^m$.
However, since they are based on the relations
between mean quantities, below they are not used
separately but are averaged for the set of stars. We
found the mean dependence of the age on dereddened
color from the ages calculated in this way:
\begin{equation}
\label{exp2}
T=0.42e^{3.86(B_{T}-V_{T})_{0}},
\end{equation}
It agrees well with the analogous dependence (3)
for the GCS ages, and these curves coincide on
the scale of Fig. 2. Although this coincidence for
$(B_{T}-V_{T})_{0}>0.45^m$ is partly explained by the use of
dependence (4) to estimate the ages, the age estimates
in the remaining range are independent, and
we see good agreement between the ages derived here
and GCS ages. Thus, although the accuracy of the
derived ages (predominantly 50--100\%) is lower than
that of the GCS ones (40--70\%) and they can be used
only in averaging for the set of stars, we see that
broad-band photometry can replace narrow-band one
in this case.

\section*{KINEMATICS}

When analyzing the distribution of stars in the
$UV$ plane, we divided the sample of 15 402 thin-disk
stars into several age subsamples. When the age
dependence of the remaining kinematic parameters
(dispersions $\sigma(U)$, $\sigma(V)$, $\sigma(W)$, Ogorodnikov--Milne
model parameters, Oort constants $A$, $B$, $C$, $K$, and
solar motion components $U_{\odot}$, $V_{\odot}$, $W_{\odot}$) is analyzed,
modern computers allow a moving calculation to be
applied instead of calculations for a few subsamples.
The stars are arranged by age and the mean age,
along with the kinematic parameters, is calculated for
1000 minimum-age stars. The minimum-age star is
then excluded from the set of stars under consideration,
a previously unused minimum-age star is then
introduced instead of it, and the calculations of the
mean age and kinematic parameters are repeated. As
a result, we obtain $15402-1000=14402$ mean ages
with the corresponding set of kinematic parameters.
The window of calculations with a width of 1000 stars
corresponds to the range of ages from 0.1 Gyr for the
youngest stars to 1 Gyr for stars with an age of 5 Gyr.

\subsection*{Age Dependence of the Velocity Dispersion}

The variations in dispersions $\sigma(U)$, $\sigma(V)$, $\sigma(W)$
with age for 15 402 thin-disk stars are indicated
in Fig. 4 by the solid black curve. The gray band along
the black curve indicates the range of errors. We see
that some of the dispersion fluctuations around the
gradual increase exceed the errors and, therefore, are
real (for a discussion, see below).

The gradual increase in dispersions is well (the
correlation coefficients are more than 0.97) fitted by
power laws, which are indicated in Fig. 4 by the
dashed lines: $\sigma(U)=22.41T^{0.307}$, $\sigma(V)=14.39T^{0.239}$, $\sigma(W)=10.26T^{0.353}$. Holmberg
et al. (2009) also found a power-law increase in
dispersions with age based on GCS data and interpreted
it as continuous kinematic ``heating'' of the
thin disk during its lifetime, for example, by the radial
migration of metal-poor stars from the inner (with
respect to the Sun) Galactic regions according to
the theory of Sch\"onrich et al. (2010). It follows from
this theory that the kinematic parameters considered
here should depend not on age but on metallicity.
The observed age--metallicity correlation does not yet
allow us to decide what argument the variations of
kinematics depend on. It is also possible that both
views are correct: the radial migrations of metalpoor
stars affect the kinematics in some periods, while
other mechanisms increasing the velocity dispersion
with age irrespective of the metallicity affect it in the
remaining time.

The stepwise polygonal curves in Fig. 4 indicate
the dependences adopted in the BMG from the results
of Gomez et al. (1997). We see that these steps are a
rather rough approximation of the gradual increase in
dispersions, but, on average, the derived $\sigma(U)$ corresponds
to the BMG, $\sigma(V)$ is appreciably smaller, and
$\sigma(W)$ for old stars is larger than that in the BMG.
Gomez et al. (1997) used $\pi$ and $\mu$ from Hipparcos,
dissimilar $V_r$, a set of isochrones, Str\"omgren photometry,
and other data that, on the whole, are inferior
to our data in accuracy and the number of stars: for
example, the limiting accuracy of $U$, $V$, and $W$ is
15 km s$^{-1}$ against 5 km s$^{-1}$, the number of stars
younger than 5 Gyr is about 1500 against 14 350.
Consequently, the inaccuracy and incompleteness of
the data by Gomez et al. is the most likely cause of
the discrepancy in dispersion estimates. In particular,
the discrepancy in $\sigma(W)$ is explained by the difference
in estimating the kinematic boundary of the thin and
thick disks: 24 km s$^{-1}$ here instead of 17.5 km s$^{-1}$
in the BMG. It can be seen from Fig. 1 that, having
adopted the lower value, we lose the fast old thin-disk
stars. Obviously, they were lost by Gomez et al.
due to the low accuracy of $V_r$, because the old stars
located mostly comparatively far from the Galactic
plane have comparatively large $V$ due to Galactic
rotation and small $W$ in this case, erroneous $V_r$ give
a large relative error of $W$ and, in combination with
large $V$, this increases the probability of assigning
an erroneously large total velocity to a star and classifying
it as a thick-disk or halo star. The opposite
discrepancy in $V$ is explained in the same way: here,
erroneous $V_r$ give a large relative error of $V$ that is
canceled out by large $V$ and the thick-disk stars with
small $\mu$, $U$ and $W$ are erroneously considered thin-disk
ones.

The individual symbols with error bars in Fig. 4
indicate the extraneous results: the gray squares --
Dehnen and Binney (1998), the result was obtained
only from $\mu$; the gray circles -- Francis and Anderson
(2009); the gray diamonds -- Torra et al. (2000);
the black triangles -- Mignard et al. (2000), the result
was obtained only from $\mu$; the black diamonds -- Chen
et al. (1997); the black squares -- Gontcharov (2011);
the black circles -- Famaey et al. (2005). All these
results are in agreement with those obtained here,
although it should be noted that many of these results
are not completely independent, because the data
from the same sources are used everywhere, while
$V_r$ with corrected systematic errors for stars over the
entire sky are used only in Gontcharov (2011) and this
study.

Famaey et al. (2005) and Gontcharov (2011) obtained
their results for red giant branch KIII and MIII
stars with presumed ages from 3 to 10 Gyr. Here,
arbitrary ages of 4.7 and 4.9 Gyr for KIII and MIII
stars, respectively, were assigned to them, and they
point out an almost constant dispersion level for stars
older than 4.5 Gyr. This absence of an increase in
dispersion for old stars at a noticeable increase for
young ones is known as the Parenago (1950) discontinuity.
It was considered in detail by Dehnen and
Binney (1998); according to them, the dispersions
are stabilized at $(B_{T}-V_{T})_{0}\approx0.61^m$, i.e., according to
Eq. (5), at an age of about 4.4 Gyr. They offered a
possible, though not conclusively proven explanation:
the color correlates with the age for O--F dwarfs and
subgiants and does not correlate for G--M, with the
boundary lying near the color for which the lifetime
of a star near the MS is equal to the lifetime of the
Galactic disk, i.e., 10 Gyr for solar-type stars. This
is because there are no old stars among the O–F
dwarfs and subgiants, while the G--M dwarfs and
subgiants are distributed in ages from 0 to 10 Gyr
rather uniformly. It is important that the color--age
correlation pointed out by Dehnen and Binney underlies
the method applied in this paper. Therefore,
in fact, the kinematics of dwarfs and subgiants bluer
than the color at which the Parenago discontinuity
manifests itself is considered here.

Dehnen and Binney (1998) and Mignard (2000)
obtained their results using only $\mu$, without $V_{r}$. Their
systematic deviation to lower values from $\sigma(U)$ and $\sigma(W)$
obtained here appears to be explained in the
same way as the smaller $\sigma(W)$ in the BMG was
explained previously: Galactic rotation dominating in
the motion of stars for the velocity components $U$ and
$W$ manifests itself mainly in $\mu$, and the remaining
motions in $U$ and $W$ are difficult to reveal against the
background of Galactic rotation without invoking $V_r$.

The vertical arrows in Fig. 4a indicate four distinct
periods of sharp $\sigma(U)$ spikes against the background
of a gradual rise: about 0.6, 1.2, 2.3, and 3.1 Gyr
ago. The same $\sigma(U)$ spikes are seen in the results
of Dehnen and Binney (1998) (gray squares), Francis
and Anderson (2009) (gray circles), and Gomez
et al. (1997) (absent in the figure). As we show below,
when the distribution of stars in the $UV$ plane is
analyzed, the sharp increase in $\sigma(U)$ is caused by the
formation or intrusion of stellar streams in the region
of space under consideration. Since no sharp increase
in dispersion manifested itself in these periods or it
was indistinct in Figs. 4b and 4c, these streams have
a predominantly radial direction (along the $X$ axis).

Figure 5 shows the age dependence for the shape
of the velocity ellipsoid, i.e., for the dispersion ratios
$\sigma(V)/\sigma(U)$ (black solid curve) and $\sigma(W)/\sigma(U)$ (gray
solid curve) in comparison with the same quantities
from the BMG (the black and grays dashed curves,
respectively). The deviation of our result from the
BMG is explained by the systematic difference between
$\sigma(V)$ and $\sigma(W)$ mentioned above. The influence
of the radial streams is also seen in these
results as the periods of a decrease in both ratios. The
overall pattern of the curves suggests a strong deviation
of the velocity distribution from the theoretically
justified ellipsoid for stars younger than 0.6 Gyr and
stabilization of these ratios for stars older than about
1.5 Gyr at $\sigma(V)/\sigma(U)\approx0.6$
and $\sigma(W)/\sigma(U)\approx0.5$.
The results by Gomez et al. (1997) show the same for
stars younger than 1.5 Gyr, but stabilization for old
stars occurs at 0.7 and 0.62, respectively. Famaey
et al. (2005) obtained, on average, about 0.65 and
0.5 for old red giant branch stars, i.e., close to the
stabilization levels of these quantities obtained here.
Their simulations gave 0.79 and 0.55, which differ
from the empirical ones due to selection in favor
of slow stars because of sample incompleteness.
Gontcharov (2011) and Bobylev et al. (2009) pointed
out the variations in these rations with coordinate $Z$.
Thus, there is a great variety of estimates, but, as
has been noted above, the accuracy and volume of
the data used are primarily important. Therefore, the
values obtained here appear most plausible.

\subsection*{Distribution of Stars in the $UV$ Plane}

The distribution of the selected thin-disk stars in
projection onto the $UV$ plane is shown in Fig. 6. In
this case, the sample was divided into the following
subsamples to reveal the streams: (a) 2244 stars
younger than 0.4 Gyr, (b) 1938 stars with ages
0.4--0.9 Gyr, (c) 3062 stars with ages 0.9--1.9 Gyr,
(d) 3332 stars with ages 1.9--2.8 Gyr, (e) 1875 stars
with ages 2.8--3.6 Gyr, and (f) 1899 stars with ages
3.6--5 Gyr. In support of the reality of the calculated
ages, the main known superclusters (dynamical
streams) appear and disappear in the graphs in complete
agreement with the theoretical and empirical
results by Antoja et al. (2008) presented in their
Fig. 13, Fig. 6a is consistent with their graph for stars
with ages 0.1--0.5 Gyr, Figs. 6b and 6c are consistent
with their graph for stars with ages 0.5--2 Gyr, etc.,
as well as with the empirical results by Francis and
Anderson (2009) presented in their Fig. 9. The
main streams are (the mean values of $U$, $V$, and $W$
were taken from Antoja et al. (2008) and Bobylev
et al. (2010)): the Pleiades ($U\approx-14$ km s$^{-1}$, $V\approx-23$ km s$^{-1}$)
becomes smaller in the number of
members than other streams in accordance with the
age of the cluster itself, about 100 Myr (Bovy and
Hogg 2010); Coma Berenices ($U\approx-11$ km s$^{-1}$, $V\approx-8$ km s$^{-1}$) becomes inferior to the Pleiades
in the number of members with age; the streams
NGC~1901 ($U\approx-25$ km s$^{-1}$, $V\approx-10$ km s$^{-1}$) and
IC~2391 ($U\approx-21$ km s$^{-1}$, $V\approx-16$ km s$^{-1}$) are
particularly clearly seen in Figs. 6a and 6b not far from
the Pleiades and Coma Berenices, being lost near the
Hyades in Figs. 6c--6f, which is in agreement with
the age of the cluster NGC~1901 itself, $0.4\pm0.1$ Gyr
(Carraro et al. 2007); the Sirius or Ursa Major stream
($U\approx+9$ km s$^{-1}$, $V\approx+3$ km s$^{-1}$) is more clearly
seen in Figs. 6b--6d at an estimated age of the cluster
itself 350--413 Myr (Bovy and Hogg 2010); the
Hyades ($U\approx-43$ km s$^{-1}$, $V\approx-20$ km s$^{-1}$) manifests
itself in Figs. 6c--6e at an estimated age of the
cluster itself $488-679$ Myr (Bovy and Hogg 2010);
the $\alpha$~Ceti or Wolf~630 stream ($U\approx+23$ km s$^{-1}$, $V\approx-28$ km s$^{-1}$)
appears in Figs. 6d--6f in accordance
with the results by Francis and Anderson (2009)
and Bobylev et al. (2010); the Hercules stream
($U\approx-30$ km s$^{-1}$, $V\approx-50$ km s$^{-1}$) is represented by
a few stars in Fig. 6c and contains an increasing
large fraction of stars when passing from Fig. 6d
to Fig. 6f; the HR~1614 stream ($U\approx+15$ km s$^{-1}$, $V\approx-60$ km s$^{-1}$)
can be seen only in Fig. 6f, i.e., for
an age of more than 3.5 Gyr ago in accordance with
the analysis by Bobylev et al. (2010).

Comparison of Figs. 6 and 4 shows that the $\sigma(U)$
spikes marked by the arrows in Fig. 4a were caused
about 0.6, 1.2, 2.3, and 3.1 Gyr ago mainly by the Sirius,
Hyades, $\alpha$~Ceti/Wolf~630, and Hercules streams,
respectively, although other streams also play some
role in all cases.

\subsection*{Solar Motion Relative to the Stars}

The solar velocity components $U_{\odot}$, $V_{\odot}$, $W_{\odot}$
are plotted against the stellar age in Fig. 7: the
black curve with the gray error band represents
our result for 15 402 thin-disk stars. This result
agrees, within the error limits, with most of the
extraneous results marked in the figure by individual
symbols: the gray squares --Dehnen and Binney
(1998), the result was obtained only from $\mu$; the
gray circles -- Francis and Anderson (2009); the gray
diamonds -- Torra et al. (2000); the black triangles --
Mignard (2000), the result was obtained only from $\mu$;
the black diamonds -- Chen et al. (1997); the black
squares -- Gontcharov (2011); the black circles --
Famaey et al. (2005); the gray triangles -- Zhu (2000);
and the black stars -- Mendez et al. (2000). We see
reliably determined smooth variations in solar motion,
with those in $W_{\odot}$ being possibly periodic, with a
period of $\sim1$ Gyr. The stable motion $2-5$ Gyr ago
changed abruptly to a stable motion in a slightly
different direction $0.9-1.7$ Gyr ago, and in the last
0.5 Gyr the motion changed most significantly over
all 5 Gyr. Naturally, not the solar motion but the
composition of the sample of stars serving as a frame
of reference or a realization of the local standard of rest
(LSR) changes. These changes result, for example,
from the appearance of the previously mentioned
radial stellar streams in the solar neighborhood under
consideration and from the deformation and displacement
of the layer of star-forming gas (for example,
in the Gould Belt in the last 0.1 Gyr, according to
Gontcharov (2009, 2012c)).

The shape of the curve in Fig. 7b reflects a smooth
variation in the revolution velocity of stars around the
Galactic center with their age -- an asymmetric drift:
the stars born later revolve more rapidly, but the Sun
overtake them as well. Many studies are devoted to
determining this peculiar solar motion. Their results
disagree mainly because the researchers refuse to
recognize the obvious dependence of the solar motion
on stellar age and because there is disagreement in
choosing ``reference'' stars for the LSR.

For example, the dashed straight line in Fig. 7b
indicates the fit to the asymmetric drift from Dehnen
and Binney (1998) for stars older than 0.5 Gyr (over
the gray squares, except the two leftmost ones). For
imaginary zero-age stars it gives $V_{\odot}=5.25$ km s$^{-1}$
widely used by researchers (more than 600 references)
and marked in the graph by the snowflake.
In reality, however, there are no zero-age stars and,
in general, no sufficient number of stars revolving so
rapidly around the Galactic center for LSR realization.
The attempts to find their analogs among the
youngest stars fail due to the peculiarity of the motion
of the latter in the Gould Belt seen in Figs. 7a and
7b: the youngest stars are not the ones revolving
most rapidly around the Galactic center. Instead of
the imaginary zero-age stars, Sch\"onrich et al. (2010)
proposed to use real stars with the maximum revolution
velocity around the Galactic center by assuming
them to be, on average, the most metal-rich ones and
by replacing the age dependence by the metallicity
dependence. These turned out to be stars with ages of
about 0.5 Gyr: the arrow and the second snowflake in
Fig. 7b mark the transfer of the Dehnen--Binney LSR
by Sch\"onrich et al. However, there is an uncertainty
here due to the membership of stars in streams as
well. It is reflected the estimated systematic error of
the result by Sch\"onrich et al. -- 2 km s$^{-1}$ at a random
error of only 0.5 km s$^{-1}$.

The failed attempts to use young stars as a realization
of the LSR forced the researchers to turn
to old red giants, but Famaey et al. (2005) detected
stellar streams here as well and directly cast doubt
on the presence of some set of stars having no net
radial motion in the solar neighborhood which can be
used as a reference against which to measure the solar
motion.

However, it can be seen from Fig. 7b that the values
of $V_{\odot}$ are nevertheless grouped near some
smooth curve that should be close to our results.
This means that we can produce a sample of stars
with a \emph{known age} relative to which the solar motion
can be determined with an accuracy, say, of about
0.2 km s$^{-1}$. The systematic motions of stars only
set some insurmountable limit for this accuracy and
make the determinations of the solar motion without
specifying the ages of the stars realizing the LSR
completely meaningless.

\subsection*{Kinematics According to the Ogorodnikov--Milne Model}

Let us determine the kinematics of the subsamples
within the linear Ogorodnikov--Milne model described by Gontcharov (2011): let us calculate the
partial derivatives of the velocity with respect to the
distance $M_{ux}, \ldots, M_{wz}$ (including the angular velocity
of Galactic rotation $\Omega_{R0}=-M_{uy}$), the Oort constants
$A$, $B$, $C$, $K$, and the vertex deviation $l_{xy}$.
The distances are an additional source of errors not
only because the accuracy of the parallaxes is limited
but also because the old stars of the sample under
consideration are rather faint and are in a very small
solar neighborhood, for example, the stars with an
age of 2 Gyr within 50 pc rather than 333 pc of the
Sun. A small range of distances gives large errors
when calculating the partial derivatives of the velocity
with respect to the distance. Therefore, let us consider
the variations in model parameters only during the
last 2 Gyr. They are shown in Fig. 8 together with
the error bands. The variations in $M_{wx}$ and $M_{wy}$ are
insignificant. All the time, except the last 0.3 Gyr, the
variations in $M_{vx}$, $M_{wz}$ and $A$ are also insignificant.
The variations in $M_{ux}$, $M_{uy}$ and $M_{uz}$, caused by the
mentioned radial streams are most significant. Sharp
transitions from one prolonged state to another are
seen.

For example, about 1.3 Gyr ago, the formation or
intrusion of the Hyades disturbed the calmness in the
solar neighborhood and changed the following quantities
for a long time: $l_{xy}$ from 0 to $+15^{\circ}$, $B$ from $-5$ to
$-20$ km s$^{-1}$ kpc$^{-1}$ (the local motions dominate over
the involvement of stars in Galactic rotation), $C$ from
0 to $-10$ km s$^{-1}$ kpc$^{-1}$, the Galactic rotation velocity
($-M_{uy}$) from $-20$  to $-40$ km s$^{-1}$ kpc$^{-1}$, $M_{vy}$ from
$-5$ to $+10$ km s$^{-1}$ kpc$^{-1}$ (the slight compression of
the set of stars along the $Y$ axis is replaced by a strong
expansion), $M_{vz}$ from 0 to $-15$ km s$^{-1}$ kpc$^{-1}$ (the
Hyades moves below the Galactic plane in the direction
of Galactic rotation), the prolonged decrease in
$M_{uz}$ is replaced by a prolonged increase (the Hyades
moves below the Galactic plane toward the Galactic
anticenter), and $M_{vx}$ becomes zero for a long time.

Here is another example. About 0.25 Gyr ago,
the processes that subsequently produced the Gould
Belt began (the last 0.18 Gyr are not considered here,
but they were analyzed by Gontcharov (2012c)): $M_{ux}$
changes from $-15$ to $+5$ km s$^{-1}$ kpc$^{-1}$ (the compression
along the $X$ axis is replaced by expansion), $M_{vz}$
and $M_{wy}$ change from $-15$ to $-20$ km s$^{-1}$ kpc$^{-1}$ and
from 0 to $+5$ km s$^{-1}$ kpc$^{-1}$, respectively, both changes
imply the emergence of rotation around the $X$ axis, $B$
changes from $-10$ to $-15$ km s$^{-1}$ kpc$^{-1}$ (the local motions increase
in importance), $C$ changes from $-10$ to $+5$ km s$^{-1}$ kpc$^{-1}$, and $K$
changes from $-8$ to 0 km s$^{-1}$ kpc$^{-1}$ (the prolonged compression
ceases).

\section*{CONCLUSIONS}

For a sample of 15 730 O--F stars with accurate
$\alpha$, $\delta$, $\mu$, and $\pi>3$ mas from the new version of
Hipparcos and $V_r$ from the PCRV, we calculated the
complete set of position and motion components $X$, $Y$, $Z$, $U$, $V$, $W$,
with the accuracy of $U$, $V$, and $W$
being higher than 5 km s$^{-1}$. The sample was divided
into 15 402 thin-disk stars, 214 thick-disk stars, and
114 halo stars. We showed that the $W$ range for
the thin-disk stars should be wider than that in the
Besancon model of the Galaxy.

The ages for the thin-disk stars independent of
$X$, $Y$, $Z$, $U$, $V$, $W$ were calculated from their
positions on the $(B_{T}-V_{T})_{0}$ -- $M_{V_{T}}$ diagram relative
to the isochrones from the Padova database for the
mean metallicity for a given color. $(B_{T}-V_{T})_{0}$ and
$M_{V_{T}}$ were calculated by taking into account the
extinction from the 3D analytical extinction model by
Gontcharov (2009) and the extinction coefficient $R_V$
from the 3D map of its variations from Gontcharov (2012a).

We analyzed the variations with age for the velocity
dispersions $\sigma(U)$, $\sigma(V)$, $\sigma(W)$, solar motion components
$U_{\odot}$, $V_{\odot}$, $W_{\odot}$, Ogorodnikov--Milne model
parameters, Oort constants $A$, $B$, $C$, $K$, vertex deviation
$l_{xy}$, and the distribution of stars in the $UV$ plane.
For all quantities, except the distribution in $UV$, we
applied a moving calculation with a window with a
width of 1000 stars.

We found smooth variations of the above kinematic
parameters with age consistent with all of the
extraneous results for which the stellar ages were
determined. The velocity dispersion variations are
well fitted by power laws the deviations from which
are explained by the short-term influence of predominantly
radial stellar streams: Sirius, the Hyades,
$\alpha$~Cet/Wolf~630, and Hercules. The derived reconciled
variations of the Ogorodnikov--Milne model parameters
and Oort constants are also caused mainly
by these streams. The accuracy of determining the
solar motion relative to the local standard of rest
was shown to be fundamentally limited due to the
variations of stellar kinematics with age.

Our study showed the possibility of using multicolor
broad-band photometry to calculate the ages
of stars and, hence, the possibility of a more thorough
analysis of the kinematics in future by invoking
not thousands of Hipparcos stars but hundreds of
thousands of Tycho-2 stars with $\alpha$, $\delta$, $\mu$, photometric
distances (derived, for example, by Gontcharov
(2011, 2012c)), which is restrained by the absence of
their $V_r$.

\section*{ACKNOWLEDGMENTS}

This study was financially supported by Program
P21 of the Presidium of the Russian Academy of
Sciences, as well as by the Federal programm of the
Ministry of education and science of the Russian Federation
``Scientific and scientific-pedagogical personnel
of innovative Russia'', stage 37, event 1.2.1.

\newpage

\begin{figure}
\includegraphics{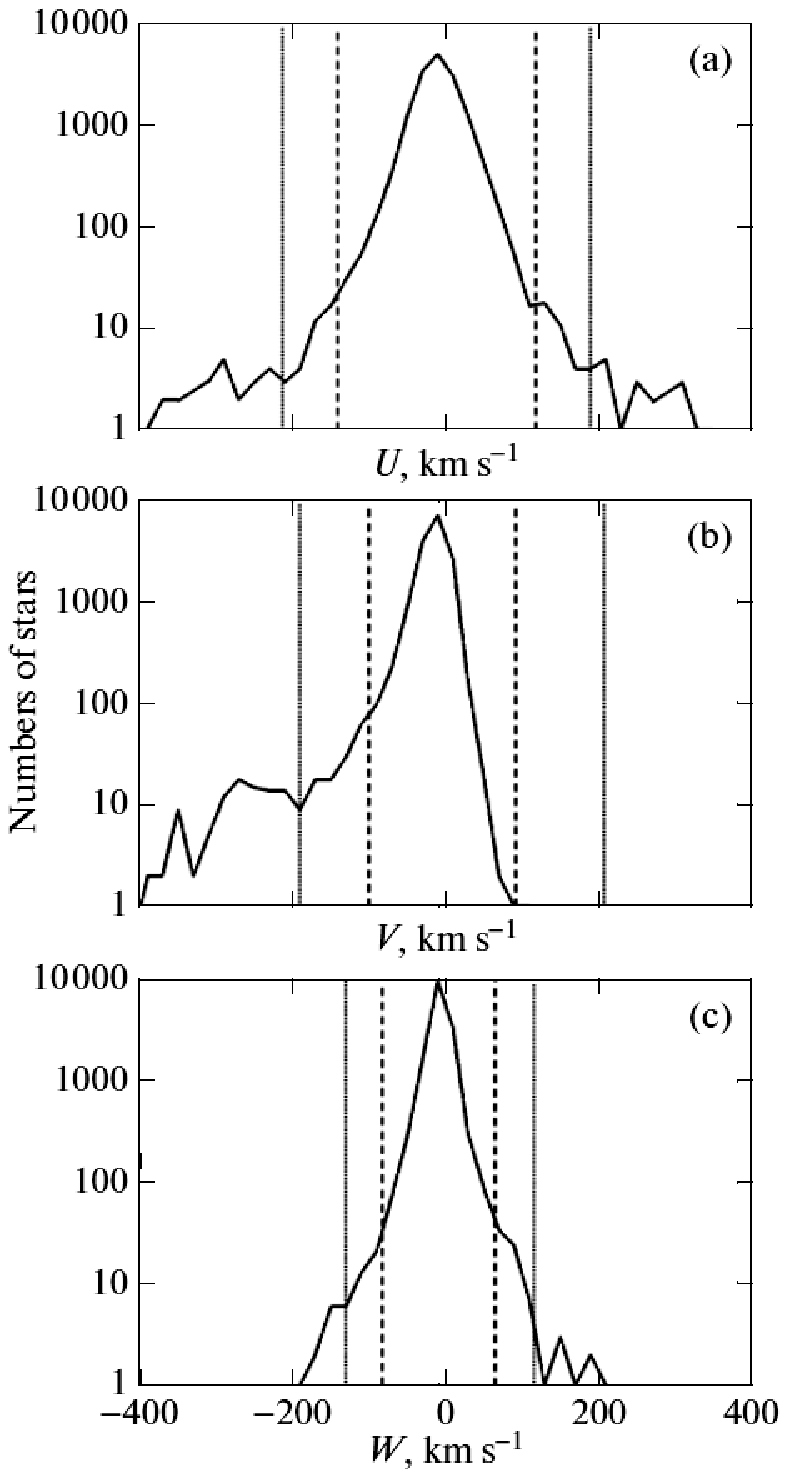}
\caption{Distribution of 15 730 stars from the initial sample
in velocity components $U$ (a), $V$ (b), and $W$ (c). The
adopted boundaries of the thin disk, the thick disk, and
the halo are indicated by the vertical dashed and dotted
straight lines, respectively.
}
\label{distr}
\end{figure}

\begin{figure}
\includegraphics{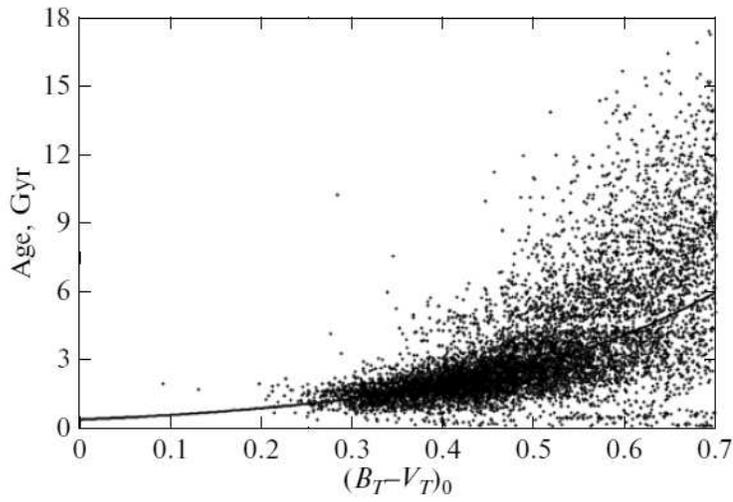}
\caption{Correlation between dereddened color $(B_{T}-V_{T})_{0}$ and age from the GCS for 8795 thin-disk stars.
The power-law fit (3) to these data is indicated by the solid curve. On the scale of the figure,
it coincides with curve (5) fitting the correlation
between $(B_{T}-V_{T})_{0}$ and age calculated here for 15 402 thin-disk stars.
}
\label{age}
\end{figure}

\begin{figure}
\includegraphics{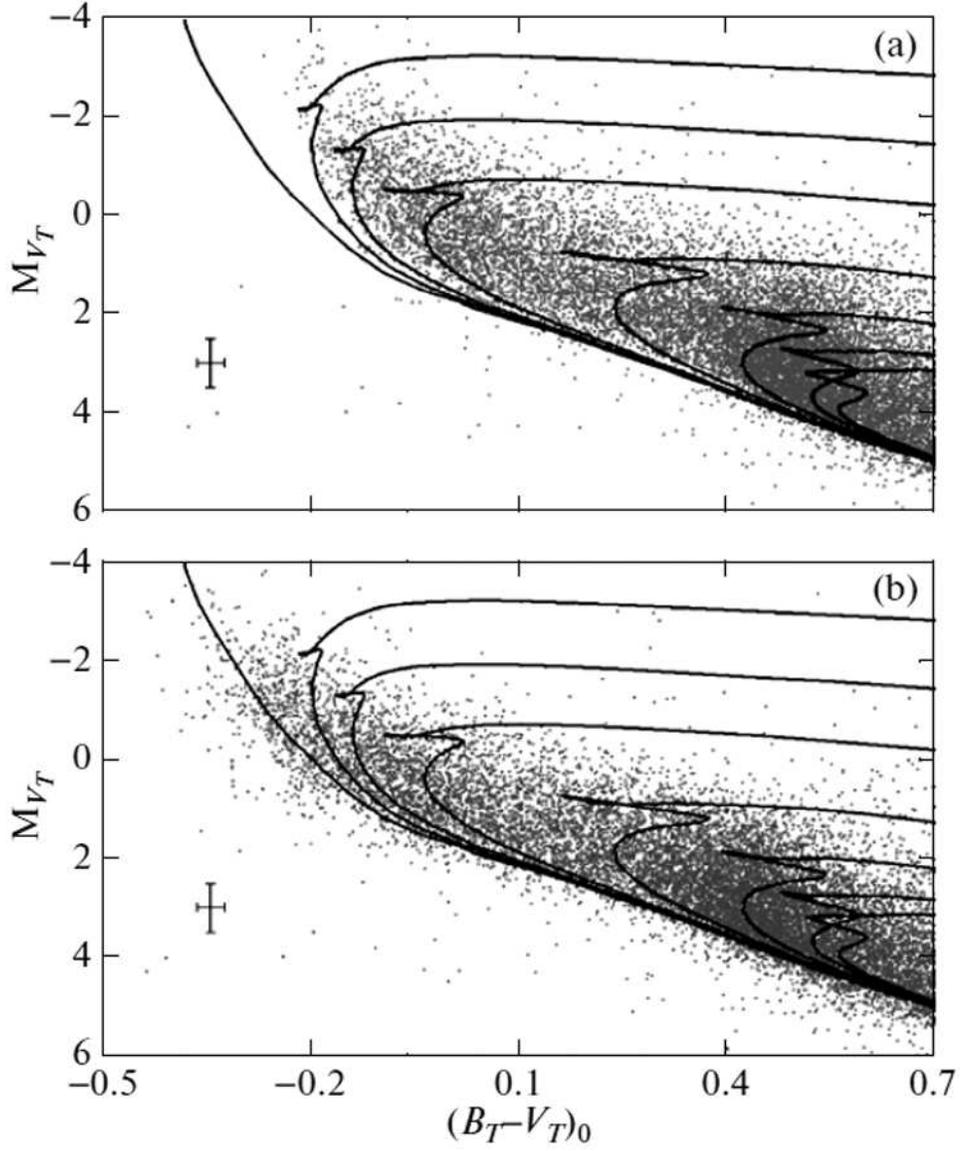}
\caption{Distribution of the 15 402 thin-disk stars under consideration on the (a) $(B_{T}-V_{T})$ -- $M_{V_{T}}$
(before dereddening) and
(b) $(B_{T}-V_{T})_{0}$ -- $M_{V_{T}}$ (after dereddening) diagrams.
The curves indicate the isochrones for solar-metallicity stars with ages
(from left to right) of 0.001, 0.1, 0.2, 0.4, 1, 2, 3, and 4 Gyr.
The cross indicates typical errors of $(B_{T}-V_{T})$ and $M_{V_{T}}$ for an
individual star.
}
\label{izo}
\end{figure}

\begin{figure}
\includegraphics{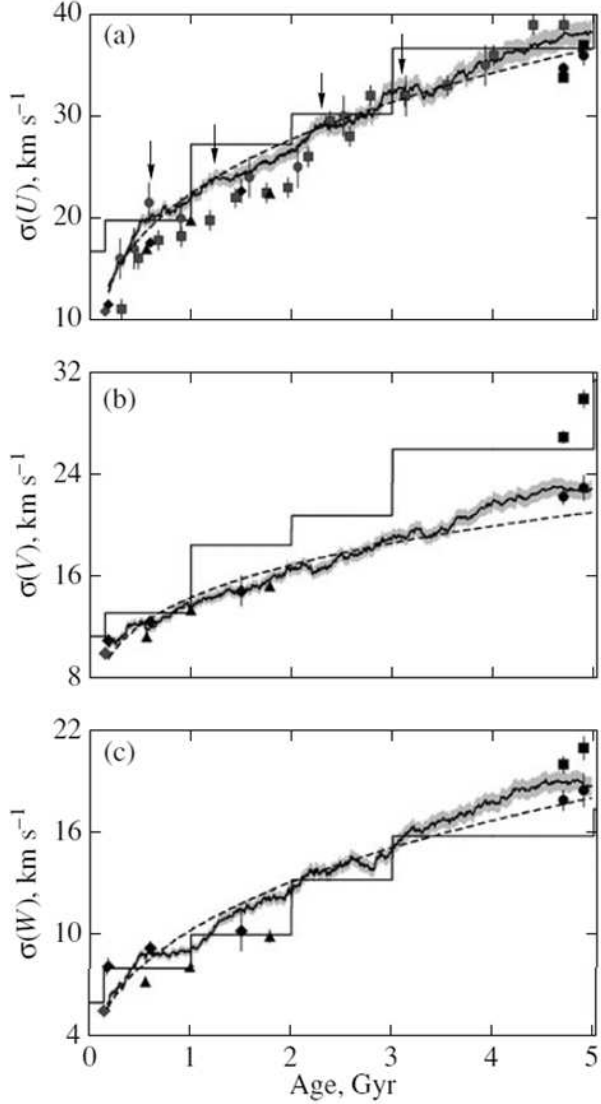}
\caption{Dispersions $\sigma(U)$ (a), $\sigma(V)$ (b), and $\sigma(W)$ (c)
versus stellar age: the black curve with a gray band of
errors indicates our result for 15 402 thin-disk stars, the
dashed curve is a power-law fit, the stepwise polygonal
curve represents the BMG, the individual symbols
represent the extraneous results listed in the text, the
arrows mark the periods of $\sigma(U)$ growth due to the stellar
streams.
}
\label{suvw}
\end{figure}

\begin{figure}
\includegraphics{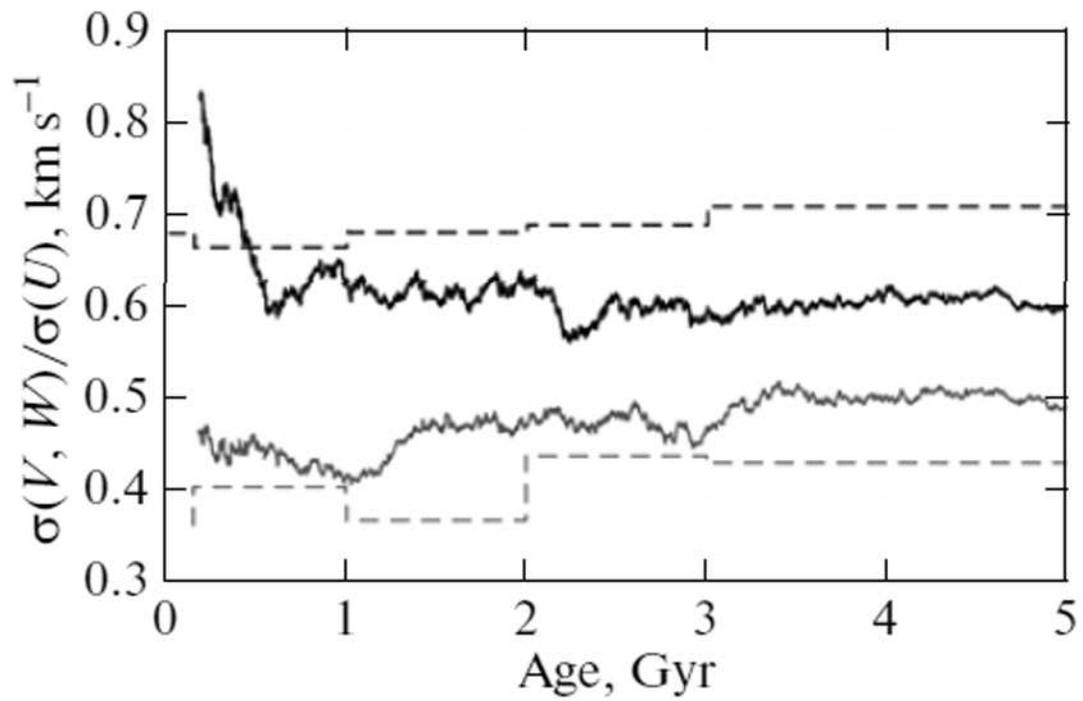}
\caption{Age dependence of the dispersions $\sigma(V)/\sigma(U)$
and $\sigma(W)/\sigma(U)$ derived here (the black and gray curves,
respectively) in comparison with the BMG (the black and
gray dashed curves, respectively).
}
\label{elli}
\end{figure}

\begin{figure}
\includegraphics{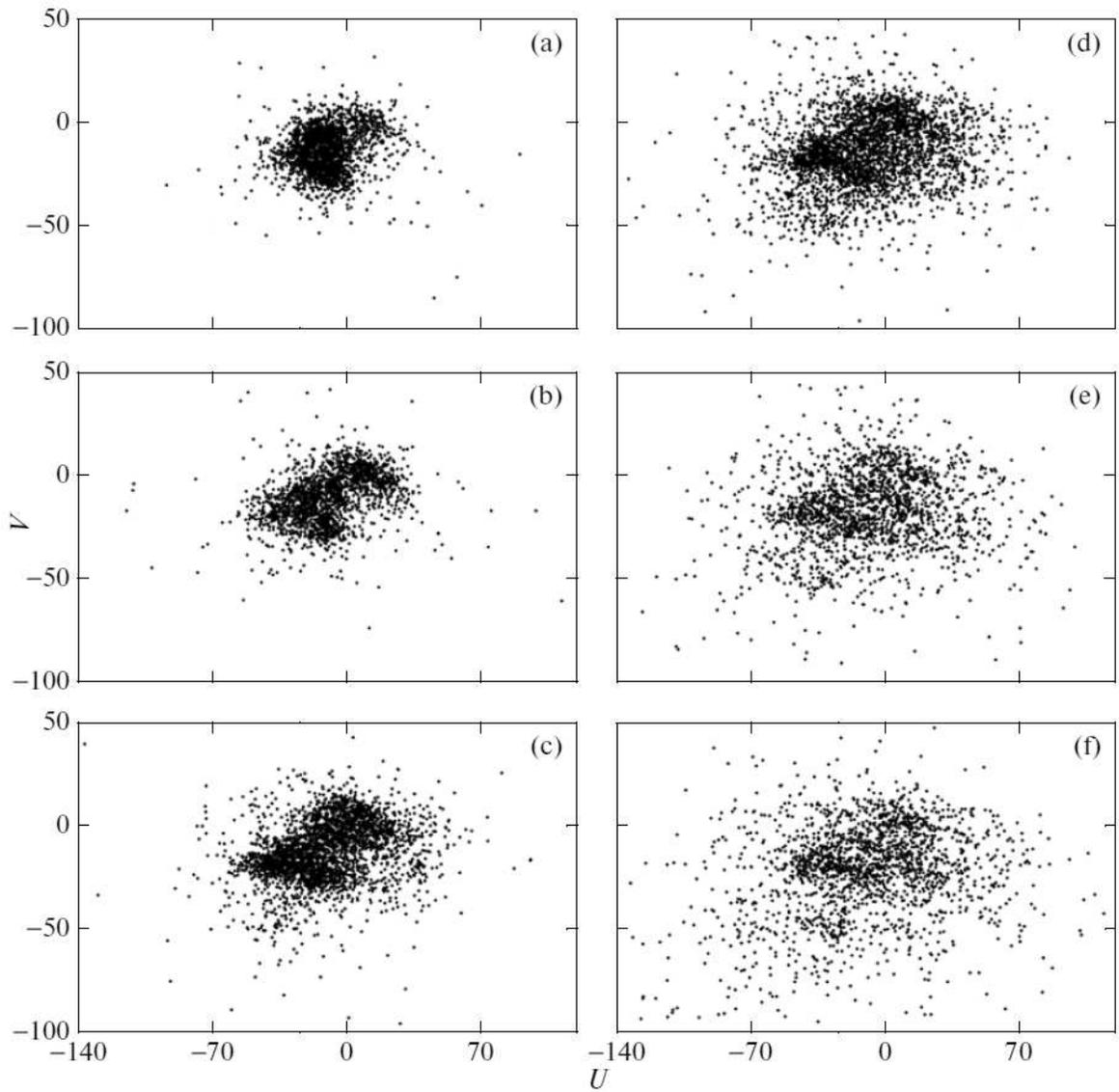}
\caption{Distribution of thin-disk stars in the $UV$ plane: (a) 2244 stars younger than 0.4 Gyr,
(b) 1938 stars with ages $0.4-0.9$ Gyr, (c) 3062 stars with ages $0.9-1.9$ Gyr,
(d) 3332 stars with ages $1.9-2.8$ Gyr, (e) 1875 stars with ages $2.8-3.6$ Gyr,
and (f) 1899 stars with ages $3.6-5$ Gyr.
}
\label{uv}
\end{figure}

\begin{figure}
\includegraphics{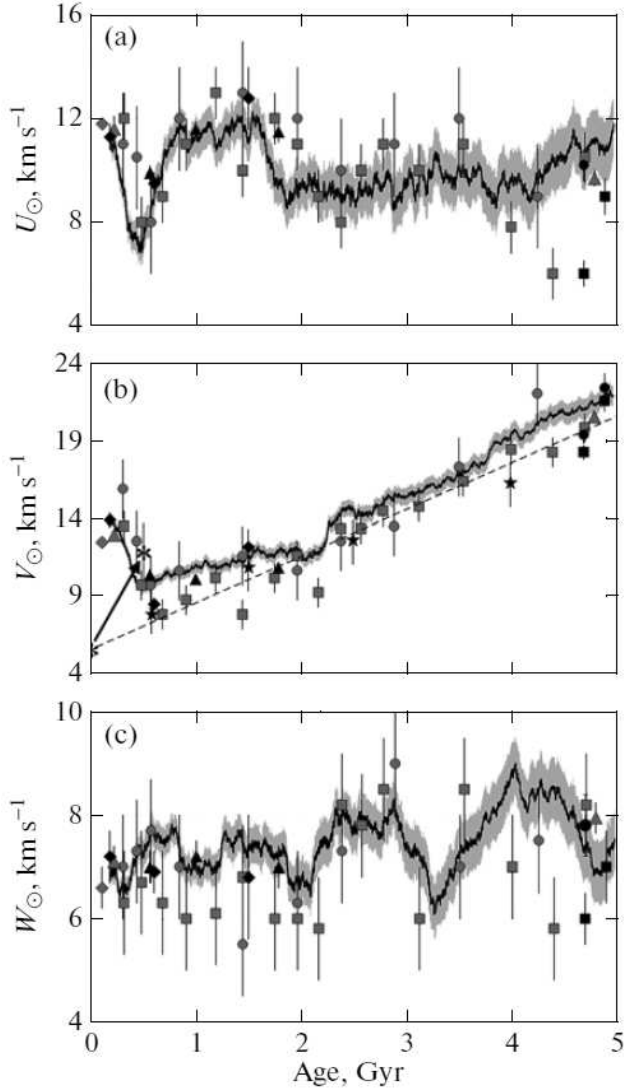}
\caption{Solar velocity components $U_{\odot}$ (a), $V_{\odot}$ (b), and
$W_{\odot}$ (c) versus stellar age: the black curve with the gray
error band represents our result for 15 402 thin-disk stars;
the individual symbols represent the extraneous results
listed in the text. The dashed straight line indicates the fit
to the asymmetric drift from Dehnen and Binney (1998),
which gives $V_{\odot}=5.25$ km s$^{-1}$ (snowflake) for their
adopted LSR; the arrow and the second snowflake mark
the transfer of this value by Sch\"onrich et al. (2010).
}
\label{uvw}
\end{figure}

\begin{figure}
\includegraphics{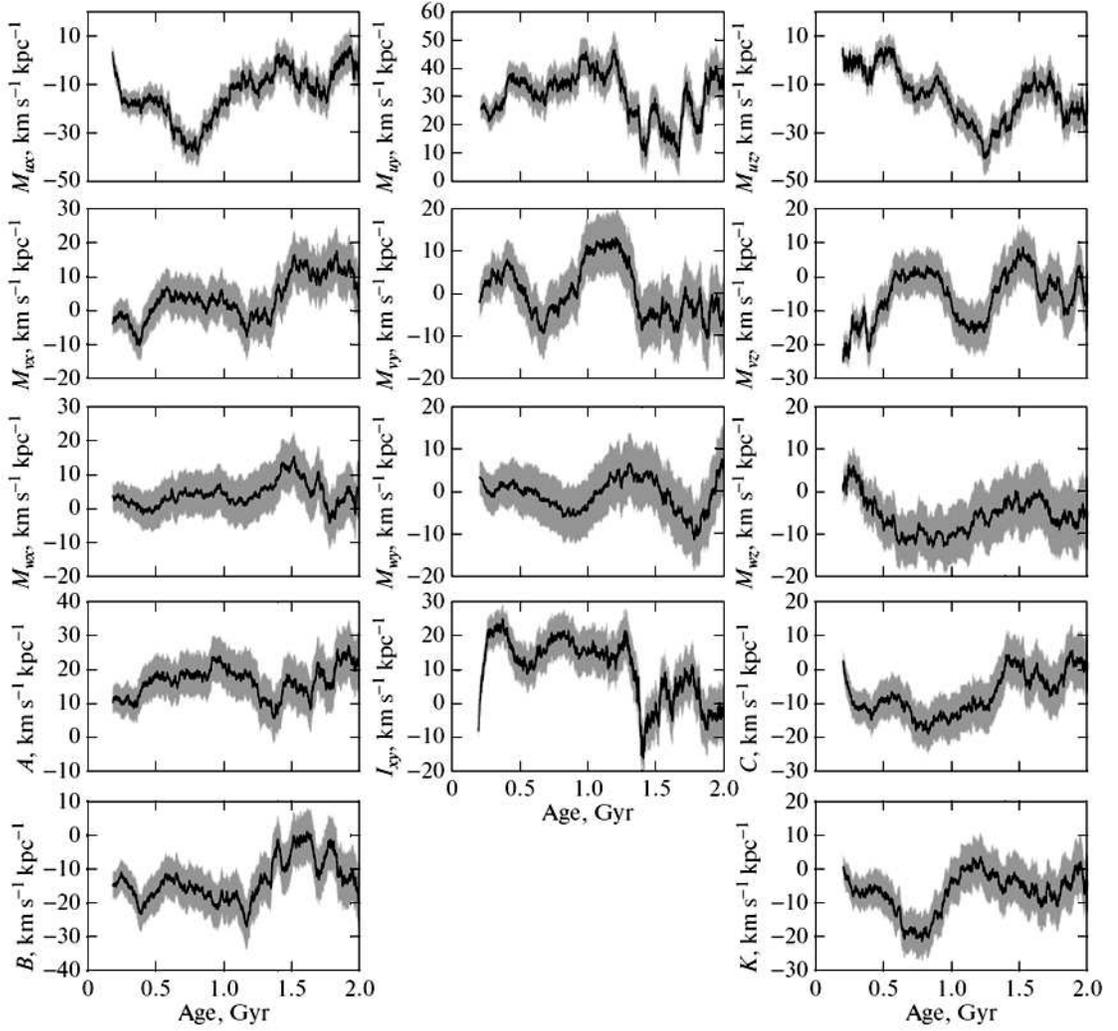}
\caption{Kinematic parameters of the Ogorodnikov--Milne model for thin-disk stars versus their age.
}
\label{abck}
\end{figure}

\end{document}